\RequirePackage{fix-cm}
\documentclass[smallextended]{svjour3}       % onecolumn (second format)
\smartqed  % flush right qed marks, e.g. at end of proof

\usepackage{mathrsfs}

\usepackage{graphicx}
\usepackage{amsmath}
\usepackage{dsfont} % it is the package of using the "\mathds{} "
\usepackage{amssymb} % it is the package of using the "\mathbb{}"%
\begin{document}
\newtheorem{assumption}{Assumption}
\newtheorem{Remark}{Remark}
\title{Estimating the Most Probable Transition Time for Stochastic Dynamical Systems%\thanks{Grants or other notes
}

\titlerunning{Estimating the Most Probable Transition Time}        % if too long for running head

\author{Yuanfei Huang         \and Ying Chao \and
          Wei Wei \and Jinqiao Duan*%etc.
}


\institute{Yuanfei Huang \and Ying Chao \and Wei Wei\at
              School of Mathematics and Statistics \& Center for Mathematical Sciences \& Hubei National Center for Applied Mathematics, Huazhong University of Science and Technology, Wuhan 430074,  China. \\
           %   Tel.: +86-17371266047\\
              \email{yfhuang@hust.edu.cn, yingchao1993@hust.edu.cn, weiw16@hust.edu.cn}           %  \\
           \and
              Jinqiao Duan \at
              Department of Applied Mathematics, Illinois Institute of Technology, Chicago, IL 60616, USA.\\
              \email{duan@iit.edu}
              \and
              $^*$Corresponding author
}

\date{Received: date / Accepted: date}

\maketitle

\begin{abstract}
This work is devoted to the investigation of the most probable transition time between metastable states for stochastic dynamical systems. Such a   system is modeled by a stochastic differential equation  with non-vanishing Brownian noise, and  is restricted in a domain with absorbing boundary. Instead of minimizing the Onsager-Machlup action functional, we examine the maximum probability that the solution process of the  system stays in a neighborhood (or a tube) of a transition path,  in order to characterize the most probable transition path. We first establish  the exponential decay lower bound  and  a power law decay upper bound for the maximum of this probability. Based on these estimates, we further derive the lower and upper bounds for the most probable transition time, under  suitable conditions. Finally, we illustrate our results in simple stochastic dynamical systems,  and highlight the relation with some relevant works.
\end{abstract}

\section{Introduction}
\label{intro}

Stochastic differential equations (SDEs) are models for  complex   phenomena in physical, chemical, biological, and  engineering systems under random fluctuations. In particular, transition phenomena between dynamically significant states, such as climate change and gene transcription, occur  under the interaction of nonlinearity and uncertainty.  It has been an interesting issue  to quantify the transition behavior between two metastable states for stochastic dynamical systems. Indeed, this has been actively  investigated  \cite{Durr1978,Kath1981,Machlup1953,Faccioli2006,Zuckerman2000,Wang2006,Gobbo2012,Langouche1982,Schulman1981,Wiegel1986,Wio2013,Khandekar2000,Huang2019,Zeitouni1987,Zeitouni1988,Chao2019}. The Onsager-Machlup method and the related Path Integrals formulation have been set up to study this problem through minimizing a  so-called Onsager-Machlup  action functional.   It is difficult to predict and describe state changes or transitions in, for example,    climate systems and Arctic sea  \cite{Field2012,Ragone2018,Moon2017}. It was reported   \cite{Ditlevsen1999,Lucarini2019} that transition phenomena in climate systems complete in finite (or even relatively short) time scales. In molecule dynamics, the transitions between two molecule states (or protein configurations)  may occur in  finite or  short time \cite{Sturzenegger2018,Hoffer2019}. In gene regulation systems,   transitions between different concentration levels of   transcription factors may happen frequently \cite{Turcotte2008,Stefan2015}.

Therefore, as a part of investigation on the transition pathways, it is crucial   to characterize the transition time.   The purpose of this paper is to estimate the most probable transition time.

We consider the following SDE on a   domain $D$ in Euclidean space $\mathbf{R}^k$:
\makeatletter
\@addtoreset{equation}{section}
\makeatother
\renewcommand{\theequation}{\arabic{section}.\arabic{equation}}
\begin{equation}\label{firstequation}
dX_t=b(X_t)dt+c \; dB_t,  \;\; ~t\geq 0.
\end{equation}
The domain  $D$ is open, bounded, connected  and is taken to have absorbing boundary.  Here $B_t$ is a standard Brownian motion in $\mathbf{R}^k$. The noise intensity $c$ is a positive constant. We assume that the drift term $b(x)$ is  in the   space $C^2(\mathbf{R}^k)$ of functions having all continuous derivatives of order up to 2. Let $x_0$ and $x_f$  be  two   distinct metastable states of the system $(\ref{firstequation})$: $b(x_0)=0$ and $b(x_f)=0$.

% ~~ X_{0}=x_0\in D.

For   system $(\ref{firstequation})$ with a given transition time $T$,  this is the usual setup for  studying transition paths between two metastable states \cite{Durr1978,Ikeda1980}:  Among all possible smooth paths connecting two metastable states ($X_0=x_0$ and $X_T=x_f$), which one is the most probable for the solution process of $(\ref{firstequation})$? The solution process of system $(\ref{firstequation})$ is almost surely nowhere differentiable. So to quantify which smooth path is the most probable one, an usual way is to compare the probabilities that the solution process stays in the neighborhood or `tube'  of such a smooth path. It was proved in \cite{Durr1978,Zeitouni1987,Zeitouni1988,Ikeda1980} that, the probability of the solution process of $(\ref{firstequation})$ in a   tube of a smooth path $\psi(t)$,  as the tube size scaling $\delta\rightarrow0$,   is
\begin{equation}\label{RN}
\mathcal{P}^{x_0}\{\|X_t-\psi(t)\|_T< \delta\}\approx\exp(-\frac{S^{OM}_T(\psi)}{c^2})\cdot \mathcal{P}^{x_0}\{\|B^c_t-x_0\|_T< \delta\},
\end{equation}
where $\mathcal{P}^{x_0}$ denotes the probability conditional on the initial position $X_0=x_0$,    $|\cdot|$ is the Euclidean norm, and $\|\cdot\|_T$ is the uniform norm:
\begin{equation}
\|\psi(t)\|_T=\sup_{0\leq t\leq T}|\psi(t)|.
\end{equation}
 Here $B^c_t=c(B_t-B_0)+x_0$ ( a shifted Brownian motion with magnitude $c$).     The Onsager-Machlup (OM) action functional is
\begin{equation}
S^{OM}_T(\psi)=\frac{1}{2}\int_{0}^{T}[\langle \dot{\psi}(t)-b(\psi(t)),(\dot{\psi}(t)-b(\psi(t)))\rangle+c^2\nabla\cdot b(\psi(t))]dt,
\end{equation}
where $\langle \cdot,\cdot\rangle$ denotes the Euclidean scalar product in $\mathbf{R}^k$. The `tube probability'  estimate  $(\ref{RN})$ shows that for a given transition time $T$, in order  to compare the probabilities that the solution process of $(\ref{firstequation})$ stays in the neighborhood of a smooth path, it can be approximately performed by comparing their corresponding OM action functionals. Thus the \textbf{most probable transition path}, when the transition time $T$ is given or known,  is defined as the one which minimizes  the OM action functional among a class of smooth paths \cite{Durr1978,Zeitouni1987,Zeitouni1988,Ikeda1980}:
\begin{equation}
\begin{split}
& \sup_{\psi\in \bar{C}_D([0,T],x_0,x_f)}\mathcal{P}^{x_0}\{\|X_t-\psi(t)\|_T< \delta\}  \\
\approx & \sup_{\psi\in \bar{C}_D([0,T],x_0,x_f)}\exp(-\frac{S^{OM}_T(\psi)}{c^2}) \cdot \mathcal{P}^{x_0}\{\|B^c_t-x_0\|_T< \delta\}     \label{dog}  \\
=& \exp\{-\frac{1}{c^2}\inf_{\psi\in \bar{C}_D([0,T],x_0,x_f)} S^{OM}_T(\psi)\} \cdot \mathcal{P}^{x_0}\{\|B^c_t-x_0\|_T< \delta\},
\end{split}
\end{equation}
where $\bar{C}_D([0,T],x_0,x_f)$ denotes the space of all absolutely continuous functions $g:[0,T]\rightarrow D$ such that $g(0)=x_0$ and $g(T)=x_f$.  As the two metastable states $x_0$ and $x_f$ are given,  we use the notation $\bar{C}_D[0,T]$ to denote $\bar{C}_D([0,T],x_0,x_f)$.   Note that when the transition time $T$ is \emph{known}, the factor  $\mathcal{P}^{x_0}\{\|B^c_t-x_0\|_T< \delta\}$  in   (\ref{dog}) does not affect the minimization.

The aforementioned  works focus on the case that the transition time $T$ is \emph{known}. However,    the transition time $T$  varies or is not known in advance in  many stochastic dynamical systems in mathematical modeling. It is thus desirable to estimate the transition time $T$.  Hence we aim to investigate the following \textbf{double optimization problem on the tube probability},  provided the tube      size $\delta$  satisfies the appropriate condition  $0<\delta<|x_f-x_0|$:
\begin{equation}\label{opt}
\sup_{T>0}\sup_{\psi\in \bar{C}_D([0,T])}\mathcal{P}^{x_0}\{\|X_t-\psi(t)\|_T< \delta\}.
\end{equation}
The \textbf{most probable transition time} for the stochastic dynamical system $(\ref{firstequation})$ transits from metastable state $x_0$ to another metastable state $x_f$ is defined as the time at which this  double optimization problem achieves its double maximum value. The corresponding  \textbf{most probable transition path} may be regarded as global.  We denote the most probable transition time as $T^{\delta}_{x_0\rightarrow x_f}$.

This paper is organized as follows. After recalling some preliminaries (Section $\ref{preliminaries}$), we investigate the probability that the solution process of the SDE stays in a neighborhood of a smooth path (Section $\ref{method}$,). We establish the exponential decay lower bound and power law decay upper bound of the maximum of  the probability for a class of transition paths. Furthermore,  we derive the bounds for the most probable transition time. Then we present two examples to illustrate our results (Section $\ref{example}$).   Finally, in Section $\ref{conclusion}$, we summarize our main results,  and discuss the connection and difference between our work and some   relevant works.

\section{Preliminaries}\label{preliminaries}

For simplicity, we consider one dimensional case (in   real line $\mathbf{R}$) from now on. We first recall some basic concepts about Brownian motion and  Wiener measure induced  by an SDE. Then we  prove results about approximation of this Wiener measure on cylinder sets, and exit properties for the   solution paths of the SDE.

\subsection{Brownian motion}
\begin{definition}
A stochastic process $\{B_t(w):t\geq0\}$ defined on a probability space $(\Omega,\mathcal{F},\mathcal{P})$ is called a Brownian motion or a Wiener process if the following conditions hold:
\begin{enumerate}
\item $B_0$=0 (a.s.);
\item The paths $t\rightarrow B_t(w)$ are continuous, a.s.;
\item $B_t$ has independent increments, i.e., if $0\leq t_1<t_2<\cdots<t_n$, then the random variables $B_{t_2}-B_{t_1},\cdots,B_{t_n}-B_{t_{n-1}}$ are independent.
\item $B_t$ has stationary increments that are Guassian distributed, i.e., $B_t(w)-B_{s}(w)$ has the normal distribution with mean 0 and variance $t-s$. Namely, $B_{t}(w)-B_s(w)\sim \mathcal{N}(0,t-s)$ for any $0\leq s<t$.
    \end{enumerate}
\end{definition}
In particular,   $c\; (B_{t}(w)-B_s(w))\sim \mathcal{N}(0,c^2(t-s))$, for  a positive constant $c$.

\subsection{Measure  induced by  the solution  process}\label{sectionp}

Let $X_t$ denote a nonexploding diffusion process on $[0,T]$ defined by the scalar stochastic differential equation in the probability space $(\Omega,\mathcal{F},\mathcal{P})$
\begin{equation}\label{diffusion1}
dX_t=b(X_t)dt+c\; dB_t,~X_0=x_0\in \mathbf{R}.
\end{equation}
The space of paths of such a diffusion process is the space $C([0,T],x_0)$ of continuous functions
\begin{equation}
C([0,T],x_0)=\{\psi(t)|\psi:[0,T]\rightarrow\mathbf{R},\psi(t)~\mbox{is continuous},\psi(0)=x_0\},
\end{equation}
with the uniform norm $\|\cdot\|_T$
\begin{equation}
\|\psi\|_T=\sup_{t\in[0,T]}|\psi(t)|,~\psi(t)\in C([0,T],x_0).
\end{equation}
In this norm,  we have the Borel field $\mathbf{B}_{[0,T]}^{x_0}$ of $C([0,T],x_0)$.

A subset $I_n$ of $C([0,T],x_0)$ in the following form is called an $n$-dimensional cylinder set:
\begin{equation}
I_n=\{\psi\in C([0,T],x_0)|(\psi(t_1),\cdots,\psi(t_n))\in H\},
\end{equation}
where $0<t_1<\cdots<t_n\leq T$ and $H$ is a Borel set  in $n$-dimensional Euclidean space. The collection of all $n$-dimensional cylinder sets is a $\sigma$-field and the class of all finite-dimensional cylinder sets is a field, which is denoted by $I$. It is known that the $\sigma$-field $\sigma(I)$, generated by $I$,  is the Borel field $\mathbf{B}_{[0,T]}^{x_0}$. That is,  $\sigma(I)=\mathbf{B}_{[0,T]}^{x_0}$.

The measure $\mu_X$ on $\mathbf{B}_{[0,T]}^{x_0}$ induced by the solution  process $X_t$ of the SDE  ($\ref{diffusion1}$) is defined by
\begin{equation} \label{Wiener}
\mu_X(B)=\mathcal{P}^{x_0}(\{w\in \Omega|~X_t(w)\in B\}),~B\in\mathbf{B}_{[0,T]}^{x_0}.
\end{equation}
Recall that such a measure induced by Brownian motion $B_t$ is the Wiener measure. For convenience, we also  call $\mu_X$ the Wiener measure induced by  solution process $X_t$.
Let $K_T(\psi,\delta)=\{x\in C([0,T],x_0)|~\|x-\psi\|_T<\delta\}$ denote the tube of a path $\psi$ with tube size $\delta$ ($i.e.$ neighborhood size),  for  $\psi\in C([0,T],x_0)$. Thus the probability that the solution process of $(\ref{diffusion1})$ stays in the $\delta$-tube of a smooth path $\psi\in \bar{C}([0,T],x_0)$ is     $\mu_X(K_T(\psi,\delta))$.

\begin{Remark}
By  the definition  (\ref{Wiener}),   the tube probability in the double optimization problem (\ref{opt}) will be estimated via  the Wiener measure $\mu_X$ for the rest of this paper.
\end{Remark}

We start with  the following lemma.
\begin{lemma}\label{lemma}
(Approximation on  cylinder sets for the measure  $\mu_X$  of a tube) \\
For a fixed transition time $T$,  a smooth path $\psi\in C([0,T],x_0)$ and a   tube size $\delta$, there exists a sequence of sets $\{J_n\}_{n=1}^{\infty}$ in the field $I$,  such that
\begin{equation}
\mu_{X}(K_T(\psi,\delta))=\lim_{n\rightarrow\infty}\mathcal{P}^{x_0}(J_n).
\end{equation}
\end{lemma}
{\bf Proof.}
Let $\mathbf{Q}$  be  the  countable set of rational numbers in  $\mathbf{R}$.   So for a fixed transition time $T>0$, the set $\mathbf{Q}\cap[0,T]$ is also countable. Define
\begin{equation}
\mathbf{Q}\cap[0,T]=\{q_0,q_1,\cdots,q_n,\cdots\}.
\end{equation}
If $T$ is a rational number, we switch the positions of 0 and $T$ with the ones of $q_0$ and $q_1$ respectly and still represent the sequence as
$\{q_0=0,q_1=T,q_2\cdots,q_n,\cdots\}$. If $T$ is an irrational number we switch the position of 0 with the one of $q_0$ and add $T$ into the sequence and denote the sequence as the same representation $\{q_0=0,q_1=T,q_2\cdots,q_n,\cdots\}$.

Introduce the subsets
\begin{equation}
\begin{split}
O_n&=\{q_0,q_1,\cdots,q_n\},\\
J_n&=\{X_t \in C([0,T],x_0)|~|X_{q}-\psi(q)|<\delta,\forall q\in O_n\}\subset I.
\end{split}
\end{equation}
Note  that $\{J_n\}_{n=1}^{\infty}$ is a decreasing sequence. Recall that the solution process of $(\ref{diffusion1})$ is almost surely continuous. So
\begin{equation}
\begin{split}
&\mu_X(K_T(\psi,\delta))\\
=&\mathcal{P}^{x_0}(\{w\in \Omega|\sup_{t\in[0,T]}|X_t(w)-\psi(t)|<\delta\})\\
=&\mathcal{P}^{x_0}(\{w\in \Omega|\sup_{t\in \mathbf{Q}\cap[0,T]}|X_t(w)-\psi(t)|<\delta\})\\
=&\mathcal{P}^{x_0}(\cap_{n=1}^{\infty}J_n)\\
=&\lim_{n\rightarrow\infty}\mathcal{P}^{x_0}(J_n).
\end{split}
\end{equation}
This completes the proof of this lemma. $\hfill\blacksquare$\\

\subsection{Non-exit probability and mean exit time of a diffusion process}\label{subsection}

In this subsection, we discuss the probability that a diffusion process $B^c_t=c(B_t-B_0)$ (where $B_t$ is a standard Brownian motion and c is a positive constant) stays in the $\delta$-tube of the origin 0 during the time period $[0,T]$. This probability was studied in   Lemma 8.1 of \cite{Ikeda1980} with unit noise intensity.  We need  to obtain the probability for different noise intensity.

Define  $G=\{x\in \mathbf{R}|~|x|<1\}$ and   $\tau^x_G(B^c_t)=\inf\{t|~B^c_t\notin G,~B^c_0=x\in G\}$. By \cite{Oksendal2003}, it is known that $\tau^x_G(B^c_t)$ is a stopping time. Then $u(t,x)=\mathcal{E}^{x}[f(B^c_t)I_{\{\tau^x_G(B^c_t)>t\}}]$ (here $\mathcal{E}^x$  denotes the expectation with respect to the probability $\mathcal{P}^x$), $x\in G$, $t>0$, is the solution of the initial value problem
\begin{equation}
\begin{cases}
\frac{\partial u}{\partial t}=\frac{1}{2}c^2\Delta u,~~u\in G,~~\\
u|_{\partial G}=0,~~u|_{t=0}=f.
\end{cases}
\end{equation}
Consequently
\begin{equation}
u(t,x)=\sum_{n=1}^{\infty}e^{-\lambda_n t}\varphi_n(x)\int_{G}\varphi_n(y)f(y)dy,
\end{equation}
where $0<\lambda_0<\lambda_1\leq \lambda_2\leq\cdots$ are eigenvalues and $\{\varphi_n(x)\}$ are corresponding eigenfunctions of the eigenvalue problem
\begin{equation}
\begin{cases}
\frac{1}{2}c^2\triangle \varphi+\lambda\varphi=0~~\mbox{in G},\\
\varphi|_{\partial G}=0.
\end{cases}
\end{equation}
The eigenfunctions (which need to be nonzero) are $\{\sin(m\pi x)\}_{m=1}^{\infty}$ and $\{\cos((m+\frac{1}{2})\pi x)\}_{m=0}^{\infty}$,  with the corresponding eigenvalues
$$\{\frac{(m\pi)^2}{2}c^2\}_{m=1}^{\infty}, \;\;\;     ~\{\frac{[(m+\frac{1}{2})\pi]^2}{2}c^2\}_{m=0}^{\infty}.$$
The set of normalized eigenfunctions form an orthonormal basis for the Hilbert space $H=L^2(-1,1)$.

In particular,
\begin{equation}\label{probability}
\begin{split}
&\mu_{B^c}(K_T(0,\delta))\\
=&\mathcal{P}^{x_0}(\{w\in \Omega|~B^c_t(w)\in K_T(0,\delta)\})\\
=&\mathcal{P}^{x_0}(\|B^c_t-0\|_T<\delta)\\
=&\sum_{n=0}^{\infty}e^{-\lambda_n T/\delta^2}\varphi_n(0)\int_{G}\varphi_n(y)dy\\
=&\sum_{n=0}^{\infty}\frac{(-1)^n4}{(2n+1)\pi}\exp\{-\frac{(2n+1)^2\pi^2c^2T}{8\delta^2}\}.
\end{split}
\end{equation}

Now we turn to discuss the mean exit time of system $(\ref{firstequation})$. The following lemma from \cite{Duan2015,Oksendal2003} is needed later.
\begin{lemma}\label{lemma2}
(Mean exit time) The mean exit time $u(x)=\mathcal{E}\tau^x_{D}(X_t)$ of the stochastic system $(\ref{firstequation})$ with $c=1$, for an orbit ($i.e.$, a trajectory) starting at $x\in D$, satisfies the following elliptic partial differential equation
\begin{equation}
Au=-1,u|_{\partial D}=0,
\end{equation}
where $\partial D$ is the boundary of $D$ and $A$ is the generator
\begin{equation}
Au=b\cdot\nabla u+\frac{1}{2}\triangle u.
\end{equation}
Moreover, if the domain $D$ has $C^{2,\gamma}$ boundary and the drift $b$ is in $C^{\gamma}(D)$ for some $\gamma\in(0,1)$, then the mean exit time $u$ uniquely exists and is in $C^{2,\gamma}(D)$.
\end{lemma}
  Recall that $C^{\gamma}(D)$ is the H\"{o}lder space consisting of functions in $D$ which are locally H\"{o}lder continuous with exponent $\gamma$. In particular,   H\"{o}lder space $C^{2,\gamma}(D)$ is the subspace of $C^2(D)$ consisting of functions whose second order derivatives are locally H\"{o}lder continuous with exponent $\gamma$. A bounded domain $D$   is called a $C^{2,\gamma}$ domain if each point of its boundary $\partial D$ has a neighborhood in which $\partial D$ is the graph of a $C^{2,\gamma}$ function. We also say that $D$ has a $C^{2,\gamma}$ boundary.

\section{Bounds for the Most Probable Transition Time} \label{method}

In this section, without loss of generality,  we take domain  $D=\{x\in \mathbf{R}|~|x-x_0|<l\}$,  with $l$   a large enough positive constant (such that $|x_f-x_0|\ll l$).  For a given tube size $\delta$, the upper and lower bounds of $\sup_{\psi\in \bar{C}_D[0,T]}\mu_X(K_T(\psi,\delta))$  are estimated  in the following two subsections. Then we   establish the bounds for the most probable transition time in Subsections 3.3.  Finally, we examine    the connection between the double optimization problems  on the tube probability  $(\ref{opt})$ and on the  OM action functional $S_T^{OM}(\psi)$, in Subsection 3.4.

%we modify the OM action functional to estimate the most probable transition time under some appropriate conditions, based on this we reformulate the double optimization problem in Subsection 3.4.

\subsection{Upper bound of $\sup_{\psi\in \bar{C}_D[0,T]}\mu_X(K_T(\psi,\delta))$   }

Define an `enlarged'  set      $D^*=\{x\in \mathbf{R}|~|x-x_0|<l+\delta\}$ and let $\tau^{x_0}_{D^*}$ be the first time that the solution process $X_t$ of system $(\ref{diffusion1})$ escapes from the domain $D^*$:
\begin{equation}
\tau^{x_0}_{D^*}=\inf\{t>0|X_0=x_0,X_t\in\partial D^*\}.
\end{equation}

For every time instant     $t>0$,
\begin{equation}
\begin{split}
\mathcal{E}\tau^{x_0}_{D^*}=&\int_{\Omega}\tau^{x_0}_{D^*}(w)d\mathcal{P}^{x_0}(w)\\
=&\int_{\{\tau^{x_0}_{D^*}(w)\leq t\}}\tau^{x_0}_{D^*}(w)d\mathcal{P}^{x_0}(w)+\int_{\{\tau^{x_0}_{D^*}(w)> t\}}\tau^{x_0}_{D^*}(w)d\mathcal{P}^{x_0}(w)\\
\geq&\int_{\{\tau^{x_0}_{D^*}(w)\leq t\}}\tau^{x_0}_{D^*}(w)d\mathcal{P}^{x_0}(w)+\int_{\{\tau^{x_0}_{D^*}(w)> t\}}td\mathcal{P}^{x_0}(w)\\
=&\int_{\{\tau^{x_0}_{D^*}(w)\leq t\}}\tau^{x_0}_{D^*}(w)d\mathcal{P}^{x_0}(w)+t\mathcal{P}^{x_0}(w|\tau^{x_0}_{D^*}(w)>t).
\end{split}
\end{equation}
According to   Lemma $\ref{lemma2}$ and the regularities of the domain $D^*$ and the drift term $b(x)$,  we know that $\mathcal{E}\tau^{x_0}_{D^*}$ is finite.    Letting $t$ tend to infinity, we obtain that
\begin{equation}
\lim_{t\rightarrow\infty}t\mathcal{P}^{x_0}(w|\tau^{x_0}_{D^*}(w)>t)=0.
\end{equation}
So for a positive constant $\epsilon$, there exists a positive constant $t_{\epsilon}$ such that for any $t>t_{\epsilon}$:
\begin{equation}
t\mathcal{P}^{x_0}(w|\tau^{x_0}_{D^*}(w)>t)<\epsilon\Rightarrow  \mathcal{P}^{x_0}(w|\tau^{x_0}_{D^*}(w)>t)<\frac{\epsilon}{t}.
\end{equation}
Notice that for every  $\psi\in \bar{C}_D([0,T])$ with  $T>t_{\epsilon}$,
\begin{equation}
\begin{split}
&\mu_X(K_T(\psi,\delta))\\
=&\mathcal{P}^{x_0}(w|~|X_t(w)-\psi(t)|<\delta,\forall t\in[0,T])\\
\leq&\mathcal{P}^{x_0}(w|~|X_t(w)-x_0|<|\psi(t)-x_0|+\delta,\forall t\in[0,T])\\
\leq&\mathcal{P}^{x_0}(w|~|X_t(w)-x_0|<l+\delta,\forall t\in[0,T])\\
=&\mathcal{P}^{x_0}(w|~X_t(w)\in D^*,\forall t\in[0,T])\\
\leq&\mathcal{P}^{x_0}(w|~\tau^{x_0}_{D^*}(w)>T)\\
<&\frac{\epsilon}{T}.
\end{split}
\end{equation}
Thus for $T>t_{\epsilon}$,
\begin{equation}
\sup_{\psi\in \bar{C}_D[0,T]}\mu_X(K_T(\psi,\delta))\leq\frac{\epsilon}{T}.
\end{equation}

\subsection{Lower bound of $\sup_{\psi\in \bar{C}_D[0,T]}\mu_X(K_T(\psi,\delta))$   }

Now we discretize $X_t$ of $(\ref{diffusion1})$ in the following way:
\begin{equation}\label{descreteversion}
\begin{cases}
X_{i}-X_{i-1}=[\kappa b(X_{i})+(1-\kappa)b(X_{i-1})]\Delta t+c(B_i-B_{i-1}),i=1,\cdots n,\\
X_0=x_0,
\end{cases}
\end{equation}
For simplicity we consider the following discretized version which has the same distribution of $(\ref{descreteversion})$:
\begin{equation}
\begin{cases}
X_{i}-X_{i-1}=[\kappa b(X_{i})+(1-\kappa)b(X_{i-1})]\Delta t+B^c_i-B^c_{i-1},i=0,1,\cdots n,\\
X_0=x_0,
\end{cases}
\end{equation}
where $(B^c_{t}(w)-B^c_s(w))\sim \mathcal{N}(0,c^2(t-s))$ and the time partition is $\Pi_n$: $0=t_0<t_1=\cdots<t_n=T$ and $\kappa\in[0,1]$. For every  $\psi\in C^2_D[0,T]:=\{\psi\in C_D[0,T]|\mbox{$\psi$ has all continuous derivatives of order no more than $2$}\}$,   and using    Lemma $\ref{lemma}$, the probability $\mu_X(K_T(\psi,\delta))$ is calculated in the following way (for every $O_n$ in Lemma $\ref{lemma}$, we reorder the elements of $O_n$ from small to large and use the new sequence for the time partition $\Pi_n$):
{\small\begin{equation}\label{pathprobability}
\begin{split}
&\mu_X(K_T(\psi,\delta))\\
=&\mu_X(\sup_{t\in [0,T]}|X_t-\psi_t|<\delta)\\
=&\lim_{n\rightarrow\infty}\mathcal{P}^{x_0}(w|\{X_{t_0},X_{t_1},\cdots,X_{t_n}\}\in \mathcal{I}_n)\\
=&\lim_{n\rightarrow\infty}\int_{\{\{B^c_{i}\}~s.t.~\{X_{t_0},X_{t_1},\cdots,X_{t_n}\}\in \mathcal{I}_n\}}\mathcal{D}^c_nB^c \exp\{-\frac{1}{2}\sum_{i=1}^{n}\frac{(B^c_i-B^c_{i-1})^2}{c^2\Delta t}\}\\
=&\lim_{n\rightarrow\infty}\int_{\{|x_i-\psi_i|<\delta\}}\mathcal{D}^c_nx\exp\{-\sum_{i=1}^{n}[\frac{(x_i-x_{i-1}-[\kappa b(x_{i})+(1-\kappa)b(x_{i-1})]\Delta t)^2}{2c^2\Delta t}]\}\\
&\cdot\prod_{i=1}^n(1+\kappa b'(x_i)\Delta t)\\
=&\lim_{n\rightarrow\infty}\int_{\{|y_i|<\delta\}}\mathcal{D}^c_ny\prod_{i=1}^n(1+\kappa b'(y_i+\psi_i)\Delta t)\\
&\cdot\exp\{-\frac{1}{2}\sum_{i=1}^{n}[\frac{(y_i-y_{i-1}+\psi_i-\psi_{i-1}-[\kappa b(y_{i}+\psi_i)+(1-\kappa)b(y_{i-1}+\psi_{i-1})]\Delta t)^2}{c^2\Delta t}]\}\\
=&\int_{|y|<\delta}\mathcal{D}^cy\exp\{-\frac{1}{2}\int_{0}^{T}[\frac{(\dot{y}+\dot{\psi}-b(y+\psi))^2}{c^2}+2\kappa b'(y+\psi)]dt\}\\
=&\int_{|y|<\delta}\mathcal{D}^cy\exp\{\frac{\mathcal{U}(y+\psi)|_{0}^{T}}{c^2}-\frac{1}{2}\int_{0}^{T}[\frac{\dot{y}^2+2\dot{y}\dot{\psi}+\dot{\psi}^2+b^2(y+\psi)}{c^2}+2\kappa b'(y+\psi)]dt\}\\
=&\exp\{-\frac{1}{2c^2}\int_{0}^{T}[(\dot{\psi}-b(\psi))^2+2\kappa c^2b'(\psi)]dt\}\int_{|y|<\delta}\mathcal{D}^cy\exp\{\frac{(\mathcal{U}(y(T)+x_f)-\mathcal{U}(x_f))}{c^2}\\
&-\int_{0}^{T}[\frac{\dot{y}^2+2\dot{y}\dot{\psi}+b^2(y+\psi)-b^2(\psi)}{2c^2}+\kappa (b'(y+\psi)-b'(\psi))]dt\},
\end{split}
\end{equation}}
where $\mathcal{I}_n=(\psi_{t_0}-\delta,\psi_{t_0}+\delta)\times(\psi_{t_1}-\delta,\psi_{t_1}+\delta)\times\cdots\times(\psi_{t_n}-\delta,\psi_{t_n}+\delta)$ and $\mathcal{D}^c_nB^c=\prod_{i=1}^n\frac{dB^c_i}{\sqrt{2\pi c^2\Delta t}}$, $\mathcal{D}^cx=\lim_{n\rightarrow\infty}\prod_{i=1}^n\frac{dx_i}{\sqrt{2\pi c^2\Delta t}}$
(the notation $\mathcal{D}_n^cx$ and $\mathcal{D}^cy$ are defined in a similar way) and
\begin{equation}
\mathcal{U}(x)=\int_{x_0}^xb(y)dy.
\end{equation}

\begin{Remark}
In $(\ref{pathprobability})$ we have used the results of  the path integrals method. For more mathematical details, such as the existence of the limitation and the links between discrete approximation and continuous functionals,       see   \cite{Chaichian2001}.
\end{Remark}

For a path $\psi\in C_D([0,T])$, denote that
\begin{equation}\label{notation}
\begin{split}
&\gamma(\psi,\delta):=\bigcup_{t\in[0,T]}\{x\in \mathbf{R}|~|x-\psi(t)|<\delta\}\subseteq\mathbf{R},\\
&h_0:=\sup_{|x-x_f|<\delta}|\mathcal{U}(x)-\mathcal{U}(x_f))|,\\
&h_1(\psi,\delta):=\sup_{x\in \gamma(\psi,\delta)}|b(x)b'(x)|,\\
&h_2(\psi,\delta):=\sup_{x\in \gamma(\psi,\delta)}|b''(x)|.
\end{split}
\end{equation}

Now we give the following estimations:
{\small\begin{equation}\label{estimation}
\begin{split}
|\int_{0}^T\dot{y}\dot{\psi}dt|=&|y(T)\dot{\psi}(T)-y(0)\dot{\psi}(0)-\int_{0}^Ty\ddot{\psi}dt|\\
=&|y(T)\dot{\psi}(T)-\int_{0}^Ty\ddot{\psi}dt|\leq|\dot{\psi}(T)|\delta+\|\ddot{\psi}\|T\delta,\\
|\int_{0}^Tb^2(y+\psi)-b^2(\psi)dt|=&|\lim_{n\rightarrow\infty}\sum_{i=1}^{n}[b^2(y_{i-1}+\psi_{i-1})-b^2(\psi_{i-1})]\Delta t|\\
=&|\lim_{n\rightarrow\infty}\sum_{i=1}^{n}[2b(\theta^1_{i-1}y_{i-1}+\psi_{i-1})b'(\theta^1_{i-1}y_{i-1}+\psi)y_{i-1}]\Delta t|\\
\leq&\lim_{n\rightarrow\infty}\sum_{i=1}^{n}|2b(\theta^1_{i-1}y_{i-1}+\psi_{i-1})b'(\theta^1_{i-1}y_{i-1}+\psi)y_{i-1}|\Delta t\\
\leq& 2h_1\delta\lim_{n\rightarrow\infty}\sum_{i=1}^{n}\Delta t\\
=&2h_1\delta T,\\
|\int_{0}^Tb'(y+\psi)-b'(\psi)dt|=&|\lim_{n\rightarrow\infty}\sum_{i=1}^{n}[b'(y_{i-1}+\psi_{i-1})-b'(\psi_{i-1})]\Delta t|\\
=&|\lim_{n\rightarrow\infty}\sum_{i=1}^{n}[b''(\theta^2_{i-1}y_{i-1}+\psi_{i-1})y_{i-1}]\Delta t|\\
\leq&\lim_{n\rightarrow\infty}\sum_{i=1}^{n}|b''(\theta^2_{i-1}y_{i-1}+\psi_{i-1})y_{i-1}|\Delta t\\
\leq& h_2\delta\lim_{n\rightarrow\infty}\sum_{i=1}^{n}\Delta t\\
=&h_2\delta T,\\
\end{split}
\end{equation}}
where $\theta^j_{i-1}\in[0,1]$ for $j=1,2$ and $i=1,2,\cdots,n$,  and  we have used the well known mean value theorem.

Substituting the estimations $(\ref{estimation})$ into $(\ref{pathprobability})$,  we obtain that
\begin{equation}
\begin{split}
\mu_X(K_T(\psi,\delta))\geq\exp\{-\frac{1}{c^2}[|\dot{\psi}(T)|\delta+h_0+(h_1+\kappa c^2h_2+\|\ddot{\psi}\|)T\delta]\}\\
\cdot\exp\{-\frac{1}{c^2}S^{\kappa,c}_T(\psi)\}\mu_{B^c}(K_T(0,\delta)),
\end{split}
\end{equation}
where $S^{\kappa,c}_T(\psi)=\int_{0}^TL^{\kappa,c}(\psi,\dot{\psi})dt=\int_{0}^T\frac{1}{2}[(\dot{\psi}-b(\psi))^2+2\kappa c^2b'(\psi)]dt$.

\begin{Remark}
When $\kappa=\frac{1}{2}$, the   functional $S^{\kappa,c}_T(\psi)$ is the OM action functional. The parameter $\kappa$ sometimes plays an important role in studying the dynamical behavior of stochastic systems. For example, to describe time-reversible dynamics, the effective action based on the symmetrical (Stratonovich's, $\kappa=\frac{1}{2}$) interpretation is applied to the system with additive noise \cite{Faccioli2006,Zuckerman2000}. In \cite{Volpe2010}, an experiment on a Brownian particle near a wall suggests that the system favors the anti-It\^{o}'s ($\kappa=1$) interpretation rather than the Stratonovich's, to ensure the Boltzmann-Gibbs distribution for the final steady state.
\end{Remark}

To give a clear and intuitive approximate form of the lower bound, we consider the following family of paths $\{\phi_T(t)\}_{T>0}$:
\begin{equation}
\phi_T(t)=\frac{x_f-x_0}{T}t+x_0,~t\in[0,T].
\end{equation}
Then we have
\begin{equation}
\begin{split}
&\sup_{\psi\in \bar{C}_D[0,T]}\mu_X(K_T(\psi,\delta))\\
\geq&\mu_X(K_T(\phi_T,\delta))\\
\geq &\exp\{-\frac{1}{c^2}[|\dot{\phi}_T(T)|\delta+h_0+(h_1(\phi_T,\delta)+\kappa c^2h_2(\phi_T,\delta)+\|\ddot{\phi}_T\|)T\delta]\}\\
&\cdot\exp\{-\frac{1}{c^2}S^{\kappa,c}_T(\phi_T,\dot{\phi}_T)\}\mu_{B^c}(K_T(0,\delta))\\
=&\exp\{-\frac{h_0}{c^2}\}\exp\{-\frac{1}{c^2}[\frac{|x_f-x_0|}{T}\delta+(h_1(\phi_T,\delta)+\kappa c^2h_2(\phi_T,\delta))T\delta]\}\\
&\cdot\exp\{-\frac{1}{c^2}S^{\kappa,c}_T(\phi_T,\dot{\phi}_T)\}\exp\{\ln(\mu_{B^c}(K_T(0,\delta)))\}\\
:=&c_0\exp\{-\frac{1}{c^2}S^{\kappa,c}_T(\phi_{T},\dot{\phi}_{T})-c_1T+\ln(\mu_{B^c}(K_T(0,\delta)))-\frac{|x_f-x_0|\delta}{Tc^2}\},
\end{split}
\end{equation}
where
\begin{equation}\label{coefficient}
\begin{split}
c_0=&\exp\{-\frac{h_0}{c^2}\},\\
c_1=&\frac{1}{c^2}[(h_1(\phi_T,\delta)+\kappa c^2h_2(\phi_T,\delta))\delta],
\end{split}
\end{equation}
are positive constants depending on the tube $K_T(\phi_T,\delta)$. Notice that all the paths $\{\phi_T(t)\}_{T>0}$ share the same curve in the state space $\mathbf{R}$ which means the two constants $c_0,c_1$ are independent of time $T$.  Here we have used the fact  that $x_f$ and $x_0$ are distinct metastable states.  Furthermore, we obtain that
{\small\begin{equation}\label{exponent}
\begin{split}
&c_0\exp\{-\frac{1}{c^2}S^{\kappa,c}_T(\phi_{T},\dot{\phi}_{T})-c_1T+\ln(\mu_{B^c}(K_T(0,\delta)))-\frac{|x_f-x_0|\delta}{Tc^2}\}\\
=&c_0\exp\{-\frac{1}{2}\int_0^T[\frac{(\dot{\phi}_T-b(\phi_T))^2}{c^2}+2\kappa b'(\phi_T)]dt-c_1T+\ln(\mu_{B^c}(K_T(0,\delta)))\\
&-\frac{|x_f-x_0|\delta}{Tc^2}\}\\
=&c_0\exp\{-\frac{1}{2}\int_0^T[\frac{((\frac{x_f-x_0}{T})-b(\phi_T))^2}{c^2}+2\kappa b'(\phi_T)]dt-c_1T+\ln(\mu_{B^c}(K_T(0,\delta)))\\
&-\frac{|x_f-x_0|\delta}{Tc^2}\}\\
=&c_0\exp\{-\frac{(x_f-x_0)^2}{2Tc^2}+\frac{(x_f-x_0)}{Tc^2}\int_0^Tb(\phi_T)dt-\frac{1}{2}\int_0^T[\frac{b^2(\phi_T)}{c^2}+2\kappa b'(\phi_T)]dt-c_1T\\
&+\ln(\mu_{B^c}(K_T(0,\delta)))-\frac{|x_f-x_0|\delta}{Tc^2}\}\\
=&c_0\exp\{-\frac{(x_f-x_0)^2}{2Tc^2}+\frac{(x_f-x_0)}{c^2}\int_0^1b(\phi_1)dt-\frac{T}{2}\int_0^1[\frac{b^2(\phi_1)}{c^2}+2\kappa b'(\phi_1)]dt-c_1T\\
&+\ln(\mu_{B^c}(K_T(0,\delta)))-\frac{|x_f-x_0|\delta}{Tc^2}\}\\
=&c_0\exp\{-\frac{(x_f-x_0)^2}{2Tc^2}+\frac{(x_f-x_0)}{c^2}\int_0^1b(\phi_1)dt-\frac{T}{2}\int_0^1\frac{b^2(\phi_1)}{c^2}dt-\frac{\kappa T (b(x_f)-b(x_0))}{x_f-x_0}\\
&-c_1T+\ln(\mu_{B^c}(K_T(0,\delta)))-\frac{|x_f-x_0|\delta}{Tc^2}\}\\
=&c_0\exp\{-\frac{(x_f-x_0)^2+2|x_f-x_0|\delta}{2Tc^2}+\frac{(x_f-x_0)}{c^2}\int_0^1b(\phi_1)dt-(\frac{1}{2}\int_0^1\frac{b^2(\phi_1)}{c^2}dt+c_1)T\\
&+\ln(\mu_{B^c}(K_T(0,\delta)))\}.
\end{split}
\end{equation}}
Here we use the fact that
\begin{equation}
\phi_T(t)=\phi_1(\frac{t}{T}),~t\in[0,T],
\end{equation}
and
\begin{equation}
\int_{0}^Tb(\phi_T(t))dt=T\int_{0}^Tb(\phi_1(\frac{t}{T}))d\frac{t}{T}=T\int_{0}^1b(\phi_1(t))dt.
\end{equation}

Notice that the integral term $\int_0^1b(\phi_1)dt$ is a constant. This lower bound can be rewritten as
\begin{equation}
k_0\exp\{-\frac{k_1}{T}-k_2T+\ln(\mu_{B^c}(K_T(0,\delta)))\},
\end{equation}
where
\begin{equation}
\begin{split}
k_0=&c_0\exp\{\frac{(x_f-x_0)}{c^2}\int_0^1b(\phi_1)dt\},\\
k_1=&\frac{(x_f-x_0)^2+2|x_f-x_0|\delta}{2c^2},\\
k_2=&(\frac{1}{2}\int_0^1\frac{b^2(\phi_1)}{c^2}dt+c_1).
\end{split}
\end{equation}

\subsection{Estimation of the most probable transition time}

We now summarise the estimation on the most probable transition time  in the following theorem.
\begin{theorem}\label{theorem1}
(Upper and lower bounds of the most probable transition time)\\
 For the stochastic dynamical system (\ref{diffusion1})   in a bounded domain $D$   with $C^{2,\gamma}$ boundary,  assume that the drift   $b$ is of $C^{2}(D)$, and $x_0$ and $x_f$ are two distinct metastable states. Then for every $\delta$ with  $0<\delta<|x_f-x_0|$, there exist   strictly positive upper and lower bounds for the most probable transition time $T^{\delta}_{x_0\rightarrow x_f}$.
\end{theorem}
{\bf Proof.} We derive the upper and lower bounds for the most probable transition time  separately.

\textbf{Step 1: Upper bound.}

 Combining the upper and lower bounds in previous sections  and  for every positive constant $\epsilon$,  we have
\begin{equation}
k_0\exp\{-\frac{k_1}{T}-k_2T+\ln(\mu_{B^c}(K_T(0,\delta)))\}\leq\sup_{\psi\in \bar{C}_D[0,T]}\mu_X(K_T(\psi,\delta))\preceq \frac{\epsilon}{T},
\end{equation}
where the notation $\preceq$ means the inequality holds when $T>t_{\epsilon}$ and $\{\phi_T\}_{T>0}$ is the family of paths which have been introduced earlier.  Recalling the infinite series representation of $\mu_{B^c}(K_T(0,\delta))$ in $(\ref{probability})$, and denoting that
\begin{equation}
\begin{split}
\mu_1(t):=\sum_{n=0}^{1}\frac{(-1)^n4}{(2n+1)\pi}\exp\{-\frac{(2n+1)^2\pi^2c^2t}{8\delta^2}\}),
\end{split}
\end{equation}
we have
\begin{equation}\label{inequality}
k_0\exp\{-\frac{k_1}{T}-k_2T+\ln(\mu_1(t))\}\leq\sup_{\psi\in \bar{C}_D[0,T]}\mu_X(K_T(\psi,\delta))\preceq \frac{\epsilon}{T}.
\end{equation}
Note that the function $\Theta(t)$ defined by
\begin{equation}
\begin{split}
\Theta(t):=k_0\exp\{-\frac{k_1}{t}-k_2t+\ln(\mu_1(t))\},
\end{split}
\end{equation}
is continuous and differentiable in $(0,\infty)$. Moreover,
\begin{equation}
\begin{split}
\frac{d\Theta(t)}{dt}=\Theta(t)[\frac{k_1}{t^2}-k_2+\frac{1}{\mu_1(t)}\frac{d\mu_1(t)}{dt}].
\end{split}
\end{equation}
It can be checked that  $\frac{d\mu_1(t)}{dt}$ is uniformly bounded for all  $t\in[0,\infty)$.

So we know that there exist constants $ T'_1$ and $ T'_2$, with $0<T'_1\leq T'_2<\infty$,  such that $\Theta(t)$ monotonically increases in $[0,T_1']$ starting at 0,  and  monotonically   decreases in $[T_2',+\infty)$ tending to 0 as $t\rightarrow\infty$. Thus the function $\Theta(t)$ achieves its maximum value for some $T'\in[T'_1,T'_2]$.

Define a function $\Xi(t)$:
\begin{equation}
\Xi(t)=\sup_{\psi\in \bar{C}_D[0,t]}\mu_X(K_t(\psi,\delta)).
\end{equation}
Since for every  path $\psi\in \bar{C}_D[0,t]$, the probability $\mu_X(K_t(\psi,\delta))$ is well defined (this has been introduced in Subsection $\ref{sectionp}$). Thus for any $t\in[0,\infty)$, the function $\Xi(t)$ is well defined and it is easy to see  that $0\leq\Xi(t)\leq1$.

Therefore
\begin{equation}
\Theta(t)\leq\Xi(t)\preceq \frac{\epsilon}{t},~t\in[0,\infty).
\end{equation}

For a given constant $\epsilon>0$, denote   $T^*=\max\{\frac{\epsilon}{\Theta(T')},t_\epsilon\}$. Then for any $t\in[T^*,\infty)$,  we have
\begin{equation}
\Xi(t)\leq\Theta(T')\leq\Xi(T').
\end{equation}
Thus the function $\Xi(t)$ achieves its maximum value in finite interval $[0,T^*]$,  and the  upper bound $T^*$  for the most probable transition path   is thus established.

\medskip

\textbf{Step 2: Lower bound.}

We now prove that the most probable transition time has a   positive lower bound.
Notice that $|x_f-x_0|>\delta$ and the solution process $X_t$ of $(\ref{diffusion1}$) is almost surely continuous. If such a  lower bound does not exist, then
\begin{equation}
\begin{split}
&\sup_{T>0}\sup_{\psi\in \bar{C}_D[0,T]}\mathcal{P}^{x_0}\{\|X_t-\psi(t)\|_T< \delta\}\\
=&\lim_{T\downarrow0}\sup_{\psi\in \bar{C}_D[0,T]}\mathcal{P}^{x_0}\{\|X_t-\psi(t)\|_T< \delta\}\\
\leq&\lim_{T\downarrow0}\mathcal{P}^{x_0}(\{w|~|X_T(w)-x_f|<\delta\})\\
=&0.
\end{split}
\end{equation}
This contradicts   the fact that the double supreme value is strictly positive. So the proof is complete.$\hfill\blacksquare$\\

\begin{figure}
  \centering
  \includegraphics[width=1\textwidth]{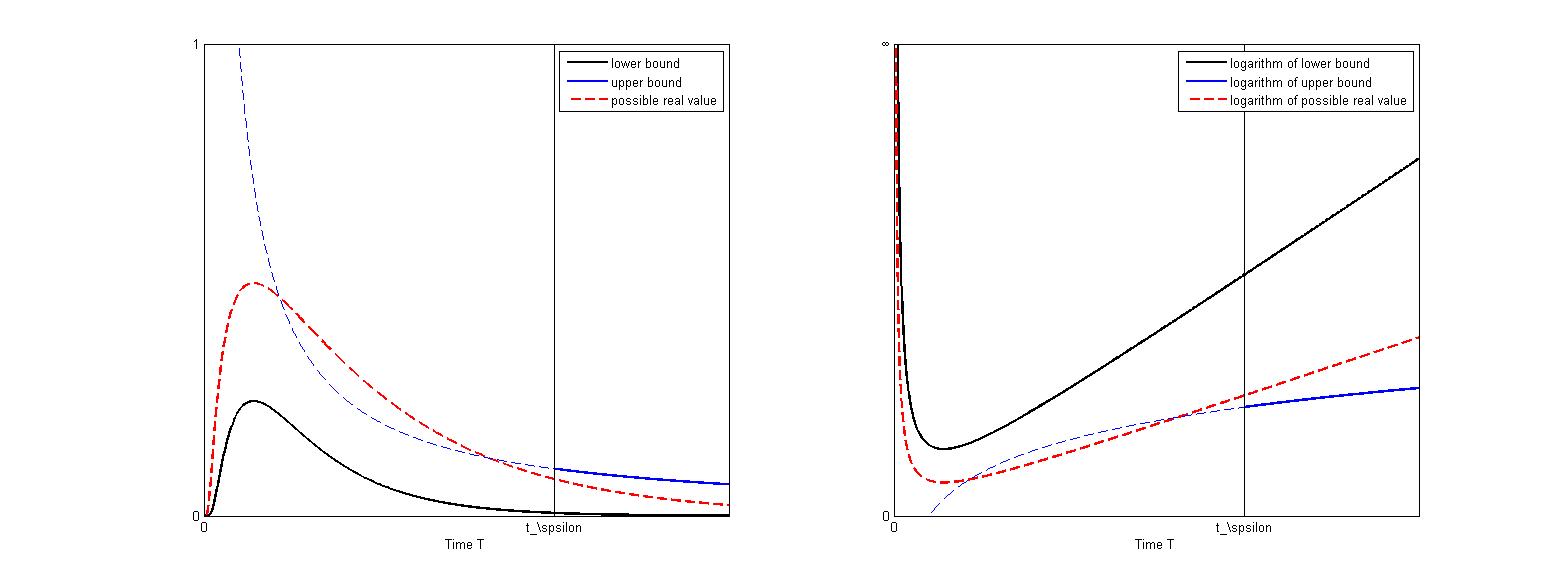}\\
  \caption{A schematic plot:  (Left)  Relationship  (\ref{inequality}).  (Right)  Logarithm of  relationship  (\ref{inequality}).}\label{demonstration}.
\end{figure}

Theorem $\ref{theorem1}$   provides   a rough range for the most probable transition time:
\begin{equation}  \label{bounds}
  0<\varrho\leq T^{\delta}_{x_0\rightarrow x_f}\leq T^*<\infty.
 \end{equation}
 Figure $\ref{demonstration}$ offers an intuitive representation of our idea. The left graph of Figure $\ref{demonstration}$ shows a schematic plot of the relationship $(\ref{inequality})$. The right graph shows the   logarithm form of the relationship.

%In fact the most probable transition time is strictly lager than 0. We state this in the following theorem.

\subsection{Modified action functional for estimating the most probable transition time      }

In this subsection, we try to find the connection between the double optimization problems on the tube probability  $(\ref{opt})$ and on the Onsager-Machlup action functional $S_T^{OM}(\psi)$.  Based on this connection, we will present an estimation  method for the most probable transition time, under   additional assumptions. To achieve this we start with the framework and assumptions of \cite{Du2020}.

We focus on a gradient system with the drift term $b(x)=-V'(x)$,  for a  potential energy $V(x)\in C^3(\mathbf{R})$, although some results can be readily extended to the non-gradient case. The path potential $U(x)$ is given by
\begin{equation}
U(x)=c^2V''(x)-\frac{1}{2}(V'(x))^2.
\end{equation}
Then $U(x)\in C^1(\mathbf{R})$ by the smoothness assumption on $V$.  We make the following assumptions on the potentials $V$ and $U$.
\begin{assumption}
There exists a local minimizer $x_m$ of $V(x)$, such that $V'(x_m)=0$, and $V''(x_m)$ is strictly positive.
\end{assumption}
\begin{assumption}
The   maximizers  of $U(x)$ are contained in a bounded domain.
\end{assumption}
\begin{assumption}
For every  $E\in\mathbf{R}$, the level set $\mathcal{L}_E=\{x\in \mathbf{R}|~U(x)=E\}$ can be decomposed into a finite number of closed and connected subsets, $i.e.$
\begin{equation}
\mathcal{L}_E=\bigcup_{k=1}^NB_k,
\end{equation}
where the subsets $B_k$ are closed and connected, and (pair-wise disjoint) $B_j\bigcap B_k=\emptyset$ if $j\neq k.$
\end{assumption}
The critical point is defined    \cite{Du2020} as follows:
 The point  $x^*\in \mathbf{R}$ is called a critical point if
\begin{equation}
U(x^*)=\max_{y\in \mathbf{R}}U(y).
\end{equation}

Denoting the set of critical points by $\Lambda$, we adopt the following   assumption.
\begin{assumption}
The set of critical points, $\Lambda$,  is discrete and has no accumulation points.
\end{assumption}

Note that
\begin{equation}
\begin{split}
S_T^{OM}(\psi)=&\int_0^T(\frac{1}{2}(\dot{\psi})^2+V'(x)\dot{\psi}-U(\psi))dt\\
=&\int_0^T(\frac{1}{2}(\dot{\psi})^2-U(\psi))dt+V(x_f)-V(x_0).\\
\end{split}
\end{equation}
We make the following assumptions on the minimizer of $S_T^{OM}(\psi)$.
\begin{assumption}\label{unique}
For    given  metastable states $x_0, x_f $ and final time $T$, the OM action functional  $S_T^{OM}(\psi)$ has a unique minimizer $\psi_T$.
\end{assumption}
\begin{assumption}\label{uniformbound}
For    given  metastable states $x_0, x_f $ and final  time $T$, there exists a positive constant $M=M(x_0,x_f)$ such that the minimizer $\psi_T$ of the  OM action functional $S_T^{OM}(\psi)$ satisfies an integral condition, i.e., the `speed' of the minimizing transition path has bounded integral:
$$\int_{0}^T|\dot{\psi}_T|dt\leq M.$$
\end{assumption}

Under   Assumption $\ref{uniformbound}$, the minimizer $\psi_T \in C^2([0,T],x_0,x_f)$ and satisfies the following Euler-Lagrangian equation:
\begin{equation}\label{ELequation}
\begin{cases}
\ddot{\psi}_T+U'(\psi_T)=0,\\
\psi_T(0)=x_0, \; \;  \psi_T(T)=x_f.
\end{cases}
\end{equation}
The proof of the smoothness of the minimizer can be found in \cite{Giaquinta2004}. It is a classical result that the energy of this Euler-Lagrangian equation is conserved along the path $\psi_T$, $i.e.$
\begin{equation}\label{conserved}
\frac{1}{2}(\dot{\psi}_T)^2+U(\psi_T)\equiv E,  \;    t\in[0,T].
\end{equation}
With   Assumption $\ref{unique}$, the value  $E$ (or denoted by $E(T)$) is uniquely determined by the initial and terminal states $x_0,x_f$ and the transition time $T$. For fixed $x_0$ and $x_f$, the value $E$ is a function of $T$ only. In this case, $E$ and $T$ are related  by the equation
\begin{equation}
T=\int_{\gamma(\psi_T)}\frac{|d\psi_T|}{\sqrt{2E-2U(\psi_T)}},
\end{equation}
where $\gamma(\psi_T)$ is the graph of $\psi_T$.

It was proved in Proposition 3 of \cite{Du2020} that for every  $T>0$,  there exists a $t_c\in[0,T]$ such that
\begin{equation}\label{ET}
U(\psi_T(t_c))+\frac{|x_0-x_f|^2}{2T^2}\leq E(T)\leq U(\psi_T(t_c))+\frac{M^2}{2T^2}.
\end{equation}
As also shown in \cite{Du2020},   $\{\psi_T\}_{T>0}$ is uniformly bounded. Since $U(x)$
and $U'(x)$ are continuous,  $\{U(\psi_T)\}_{T>0}$ and $\{U'(\psi_T)\}_{T>0}$ are also uniformly bounded.
Therefore, by further combining $(\ref{ELequation})$, $(\ref{conserved})$ and $($\ref{ET}$)$, we have the following result.

\begin{theorem}\label{dotpsi}
(Uniform boundedness for the most probable transition path)\\
 (i) The velocity for the most probable transition path is uniformly bounded after a positive time: For every positive $\varrho$,  the set of `speed' $\{\|\dot{\psi}_T\|_T\}_{T>\varrho}$ is uniformly  bounded by a constant $M_{1,\varrho}$.\\
 (ii)  The acceleration for the most probable transition path is uniformly bounded: The   set of magnitude for the acceleration $\{\|\ddot{\psi}_T\|_T\}_{T>0}$ is uniformly bounded by a constant $M_{2}$.
\end{theorem}

\medskip

Recalling the results from the  previous section,  we can further provide  an exact  lower bound of $\sup_{\psi\in \bar{C}_D[0,T]}\mu_X(K_T(\psi,\delta))$ by introducing a family of paths $\{\phi_T\}_{T>0}$.  We use the property that $\{\phi_T\}_{T>0}$, $\{\dot{\phi}_T\}_{T>0}$ and $\{\ddot{\phi}_T\}_{T>0}$ (in fact $\ddot{\phi}_T\equiv0$) are uniformly bounded. Thus, according to Theorem $\ref{theorem1}$ and Theorem $\ref{dotpsi}$,   we   give a lower bound  by  replacing $\{\phi_T\}_{T>\varrho}$ by $\{\psi_T\}_{T>\varrho}$ for some $\varrho>0$:
\begin{equation}\label{boundpsi}
\begin{split}
&\sup_{T>0}\sup_{\psi\in \bar{C}_D[0,T]}\mathcal{P}^{x_0}\{\|X_t-\psi(t)\|_T< \delta\}\\
=&\sup_{T>\varrho}\sup_{\psi\in \bar{C}_D[0,T]}\mathcal{P}^{x_0}\{\|X_t-\psi(t)\|_T< \delta\}\\
\geq&\sup_{T>\varrho}\mathcal{P}^{x_0}\{\|X_t-\psi_T(t)\|_T< \delta\}\\
\geq&\sup_{T>\varrho}\tilde{c}_0\exp\{-M_{1,\varrho}\}\exp\{-\frac{1}{c^2}S^{OM}_T(\psi_{T})-\tilde{c}_1T+\ln(\mu_{B^c}(K_t(0,\delta)))\}.
\end{split}
\end{equation}
where the coefficients $\tilde{c}_0,\tilde{c}_1$ can be determined in a similar way like $c_0,c_1$ in $(\ref{estimation})$ and ($\ref{coefficient}$).

Recall from  references \cite{Durr1978,Zeitouni1987,Zeitouni1988,Ikeda1980}, the estimation of probability $\mu_X(K_T(\psi,\delta))$ is
\begin{equation}\label{muestimation}
\begin{split}
\mu_X(K_T(\psi,\delta))\sim&\exp\{-\frac{1}{2}\int_{0}^{T}[\frac{(\dot{\psi}-b(\psi))^2}{c^2}+b'(\psi)]dt\}\mu_{B^c}(K_T(0,\delta))\\
\sim&\frac{4}{\pi}\exp\{-\frac{1}{2}\int_{0}^{T}[\frac{(\dot{\psi}-b(\psi))^2}{c^2}+b'(\psi)]dt-\frac{\pi^2 c^2T}{8\delta^2}\},~\delta\downarrow0.
\end{split}
\end{equation}
This approximation has the similar form with the lower bound in $(\ref{boundpsi})$. Note that $\frac{4}{\pi}\exp\{-\frac{\pi^2 c^2T}{8\delta^2}\}$ is the first term of the infinite series representation of the probability $\mu_{B^c}(K_T(0,\delta))$ in $(\ref{probability})$.

Moreover,  for a fixed constant $\varrho>0$ and when $T>\varrho$, we have
\begin{equation}
\begin{split}
S_T^{OM}(\psi_T)=&\int_0^T(\frac{1}{2}(\dot{\psi}_T)^2-U(\psi_T))dt+V(x_f)-V(x_0)\\
=&\int_0^T(\dot{\psi}_T)^2dt+V(x_f)-V(x_0)-E(T)T.
\end{split}
\end{equation}
It was proved in   Lemma 2 of \cite{Du2020} that
\begin{equation}
\liminf_{T\rightarrow+\infty}E(T)>0.
\end{equation}
Hence,  if
\begin{equation}\label{condition}
\frac{\pi^2 c^4}{8\delta^2}>\limsup_{T\rightarrow+\infty}E(T),
\end{equation}
then for any $\epsilon>0$ and when $T$ is large enough we have
\begin{equation}
\begin{split}
&\sup_{\psi\in \bar{C}[0,T]}\frac{4}{\pi}\exp\{-\frac{1}{c^2}S_T^{OM}(\psi)-\frac{\pi^2 c^2T}{8\delta^2}\}\\
=&\frac{4}{\pi}\exp\{-\frac{1}{c^2}S_T^{sOM}(\psi_T)-\frac{V(x_f)-V(x_0)}{c^2}-\frac{\pi^2 c^2T}{8\delta^2}\}\\
\leq& \frac{\epsilon}{T},
\end{split}
\end{equation}
and the following inequality holds for every $T>0$:
\begin{equation}
\begin{split}
&k_0\exp\{-\frac{k_1}{T}-k_2T+\ln(\mu_{B^c}(K_T(0,\delta)))\}\\
\leq&\frac{4}{\pi}\exp\{-\frac{1}{c^2}S_T^{OM}(\phi_T)-\frac{\pi^2 c^2T}{8\delta^2}\}\\
\leq&\sup_{\psi\in \bar{C}[0,T]}\frac{4}{\pi}\exp\{-\frac{1}{c^2}S_T^{OM}(\psi)-\frac{\pi^2 c^2T}{8\delta^2}\}.
\end{split}
\end{equation}
That is
\begin{equation}
k_0\exp\{-\frac{k_1}{T}-k_2T\}\mu_{B^c}(K_T(0,\delta))\leq\sup_{\psi\in \bar{C}[0,T]}\frac{4}{\pi}\exp\{-\frac{S_T^{OM}(\psi)}{c^2}-\frac{\pi^2 c^2T}{8\delta^2}\}\preceq\frac{\epsilon}{T}.
\end{equation}
This implies  that although the accuracy of the estimation $(\ref{muestimation})$ is different when $\psi$ is different,   the upper bound for this estimation,  for all paths $\psi\in\bar{C}[0,T]$, is  controlled uniformly by $\frac{\epsilon}{T}$. So the error of the estimation is also uniformly controlled.

This inspires us to use the estimation $(\ref{muestimation})$ to approximately calculate the probability of the solution process $X_t$ staying in the neighborhood of a transition path (although this estimate is quite rough for a fixed $\delta$):
\begin{equation}
\begin{split}
\mathcal{P}^{x_0}\{\|X_t-\psi(t)\|< \delta\}
\approx\frac{4}{\pi}\exp\{-\frac{1}{2}\int_{0}^{T}[\frac{(\dot{\psi}-b(\psi))^2}{c^2}+b'(\psi)]dt-\frac{\pi^2 c^2T}{8\delta^2}\}.
\end{split}
\end{equation}
We now use it to find the most probable transition time.   Define a  modified Lagrangian functional $(L^{mOM})$  by
\begin{equation}\label{lagrangian}
L^{mOM}(\psi):=\frac{1}{2}[(\dot{\psi}-b(\psi))^2+c^2b'(\psi)+\frac{\pi^2 c^4}{4\delta^2}]
\end{equation}
and the corresponding modified action functional is
\begin{equation}
S^{mOM}_T(\psi)=\frac{1}{2}\int_{0}^{T}[(\dot{\psi}-b(\psi))^2+c^2b'(\psi)+\frac{\pi^2 c^4}{4\delta^2}]dt.
\end{equation}

Thus the double optimization problem  on the tube probability  $(\ref{opt})$  in this case   is approximately equivalent to the following {\bf double optimization problem  on the modified Onsager-Machlup action functional}
\begin{equation}
\begin{split}
\inf_{T>0}\inf_{\psi\in \bar{C}[0,T]} S^{mOM}_T(\psi).   \label{newdouble}
\end{split}
\end{equation}

\begin{Remark}
We should notice that the original OM action functional comes from the path density functions, while  the modified OM action functional is derived from the   estimation of the probability that the diffusion process stays in a tube surrounding the transition path.
\end{Remark}

\begin{Remark}
The condition $(\ref{condition})$  is indeed valid for some noise intensity $c$ and tube size $\delta$. This is shown as follows,

Recall the inequalities $(\ref{ET})$:
\begin{equation}
U(\psi_T(t_c))+\frac{|x_0-x_f|^2}{2T^2}\leq E(T)\leq U(\psi_T(t_c))+\frac{M^2}{2T^2},
\end{equation}
thus
\begin{equation}
E(T)\leq c^2V''(\psi_T(t_c))-\frac{1}{2}(V'(\psi_T(t_c)))^2+\frac{M^2}{2T^2}.
\end{equation}
Since $\{\psi_T\}_{T>0}$ are uniformly bounded and the potential $V(x)$ is smooth enough, thus when $T$ is large enough, there exist some $c$ and $\delta$ such that
\begin{equation}
E(T)\leq c^2V''(\psi_T(t_c))-\frac{1}{2}(V'(\psi_T(t_c)))^2+\frac{M^2}{2T^2}< \frac{\pi^2 c^4}{8\delta^2}.
\end{equation}

Furthermore we have
\begin{equation}
\begin{split}
&\inf_{T>0}\inf_{\psi\in \bar{C}[0,T]} S^{mOM}_T(\psi)=\inf_{T>0} S^{mOM}_T(\psi_T)=\inf_{T>0} (S^{OM}_T(\psi_T)+\frac{\pi^2 c^4T}{8\delta^2}),
\end{split}
\end{equation}
and
\begin{equation}
\begin{split}
\frac{d}{dT}(S^{OM}_T(\psi_T)+\frac{\pi^2 c^4T}{8\delta^2})=\frac{dS^{OM}_T(\psi_T)}{dT}+\frac{\pi^2 c^4}{8\delta^2}=-E(T)+\frac{\pi^2 c^4}{8\delta^2},
\end{split}
\end{equation}
here the relation $\frac{dS^{OM}_T(\psi_T)}{dT}=-E(T)$ is a classical result in Hamilton-Jacobi theory whose proof can be found in \cite{Du2020,Giaquinta2004}. Thus if $E(T)$ is monotonic, then $T_{x_0\rightarrow x_f}^\delta=E^{-1}(\frac{\pi^2 c^4T}{8\delta^2})$. This implies that the global most probable transition path roughly lies on the energy shell $E=\frac{\pi^2 c^4T}{8\delta^2}$ in phase space.
\end{Remark}

\section{Examples}\label{example}
In this section we present two examples to illustrate our results.
\begin{example}
One-dimensional Brownian motion
\end{example}
\noindent Consider a scalar SDE without drift:
\begin{equation}
dX_t=c\; dB_t,~0\leq t\leq T,
\end{equation}
where $c$ is a positive constant. The modified Lagrange function is
\begin{equation}
L^{mOM}(\dot{\psi},\psi)=\frac{1}{2}(\dot{\psi})^2+\frac{\pi^2 c^4}{8\delta^2}.
\end{equation}
and the modified OM action functional is
\begin{equation}
S^{mOM}_T(\psi_T)=[\frac{x_f-x_0}{T}]^2\frac{T}{2}+\frac{\pi^2 c^4T}{8\delta^2}.
\end{equation}
So by minimizing this functional (setting its first derivative with respect time $T$ to be zero), we obtain  the estimation for the most probable transition time $T^{\delta}_{x_0\rightarrow x_f}$
\begin{equation}
T^{\delta}_{x_0\rightarrow x_f}=\frac{2\delta|x_f-x_0|}{\pi c^2}.
\end{equation}

\begin{figure}
  \centering
  \includegraphics[width=1\textwidth]{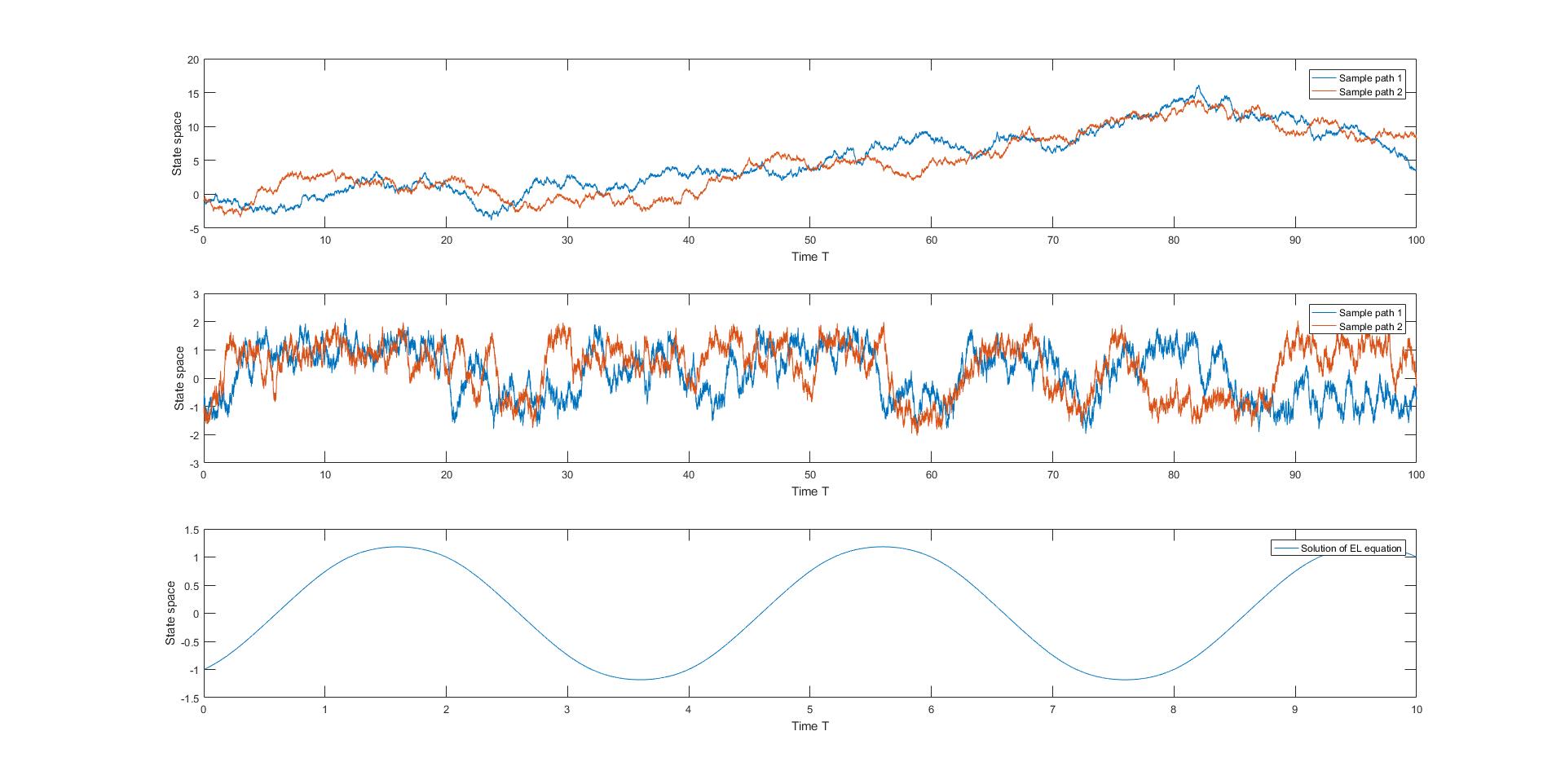}\\
  \caption{Top: Sample paths of a Brownian motion. Middle: Sample paths of the stochastic double well system (\ref{doublewellSDE}).  Bottom: The solution of the Euler-Lagrange equation for  the stochastic double well system   by a shooting method,  with $x_0=-1,x_f=1$ and $T=10.$}\label{doublewell1}
\end{figure}

\begin{figure}
\centering
\includegraphics[width=0.7\textwidth]{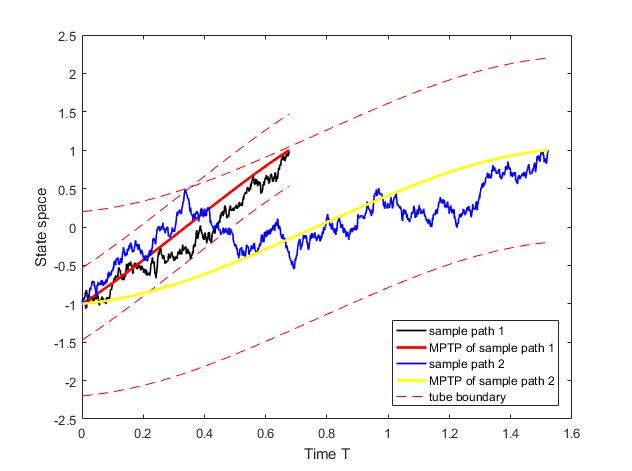}\\
\caption{Two sample transition paths of the stochastic double well system  (\ref{doublewellSDE}). Black: The transition time of sample path 1 is 0.676 and the tube size is $\delta=0.47$; Blue:  the transition time of sample path 2 is 1.522 and the tube size is  $\delta=1.20$.}\label{doublewell2}
\end{figure}

\begin{example}
A stochastic double well system
\end{example}
\noindent Consider a nonlinear scalar SDE:
\begin{equation}\label{doublewellSDE}
dX_t=(X_t-X_t^3)dt+cdB_t,~0\leq t\leq T,
\end{equation}
with $c$ is a positive constant (without loss of generality we set $c=1$). The corresponding undisturbed system has three equilibrium points: -1, 0, 1 ( we know that -1 and 1 are stable equilibrium points,  and 0 is an unstable equilibrium point). In Figure $\ref{doublewell1}$, the middle graph shows two sample paths of system $(\ref{doublewellSDE})$ with initial position $x_0=-1$.   The top graph of Figure $\ref{doublewell1}$ shows two sample paths of Brownian motion   staring at 0 as a contrast. From the comparison of these two graphs it is easy to see the difference of the behavior of a diffusion process with different drift terms. The diffusion process $(\ref{doublewellSDE})$ fluctuates between two metastable states $-1$ and $1$. The bottom graph of Figure $\ref{doublewell1}$ shows the most probable transition path  (in fact it is a local minimizer of the OM action functional) calculated by a shooting method with  $T=10$. In Figure $\ref{doublewell2}$ there are two sample transition paths of this stochastic double well system. The red and yellow curves are the corresponding most probable transition paths (MPTPs), and the dash curves are the boundaries of the transition tube.

The corresponding modified Lagrangian $L^{mOM}$ is
\begin{equation}
L^{mOM}(\dot{\psi},\psi)= \frac{1}{2}[(\dot{\psi}-\psi+\psi^3)^2+c^2(1-3\psi^2)+\frac{\pi^2 c^4}{4\delta^2}],
\end{equation}
and the modified OM action functional is
\begin{equation}
S^{mOM}_{T}(\psi)=\frac{1}{2}\int_{0}^{T}[(\dot{\psi}-\psi+\psi^3)^2+c^2(1-3\psi^2)+\frac{\pi^2 c^4}{4\delta^2}]ds.
\end{equation}

We  use  Euler method to generate  sample solution paths with time step size $\Delta t=10^{-4}$. To solve the optimization problem,   we assume that the minimizer is twice differentiable and thus we obtain the Euler-Lagrange equation:
\begin{equation}
\frac{d}{dt}\frac{\partial L^{mOM}(\dot{\psi},\psi)}{\partial \dot{\psi}}=\frac{\partial L^{mOM}(\dot{\psi},\psi)}{\partial \psi}.
\end{equation}
Hence the optimization problem turns into a second order ordinary differential equation with two boundary values,  which can be solved numerically by a shooting method:
\begin{equation}
\begin{cases}
\ddot{\psi}=b'(\psi)b(\psi)+\frac{c^2}{2}b''(\psi),\\
\psi(0)=x_0, \;\;  \psi(T)=x_f.
\end{cases}
\end{equation}
We should note that the Euler-Lagrange equation determines the local minimizer (if the joint mapping $(\dot{\psi},\psi)\mapsto OM(\dot{\psi},\psi)$ is convex then the local minimizer actually is a global minimizer which has been shown in \cite{Chao2019}). A shortcoming
 of   the shooting method is restricted by the selection of the time $T$. For instance in this example, if we choose $T$=3, the shooting method fails to find the local minimizer.

\begin{figure}
 \centering
 %label{fig:subfig:a}%
 \includegraphics[width=8cm,height=6cm]{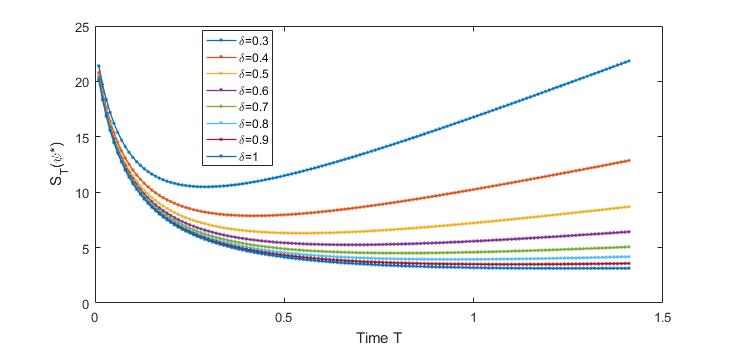}
 \caption{ The graphs of $S^{mOM}_{T}(\psi_T)$ with different tube sizes $\delta$.}\label{doublewell3}
 %\label{fig:subfig}%
\end{figure}

\begin{figure}
 \centering
 %label{fig:subfig:a}%
 \includegraphics[width=8cm,height=6cm]{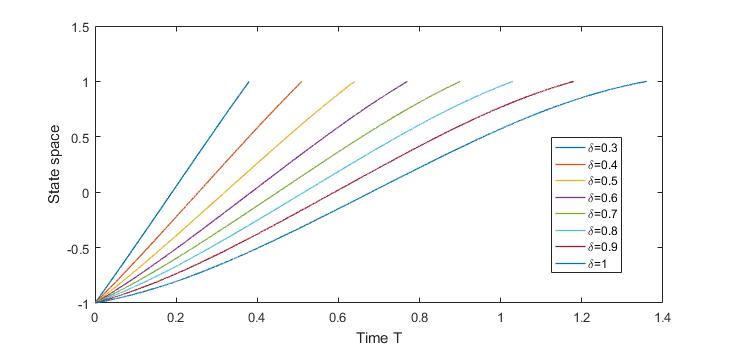}
 \caption{ The `most probable transition paths' with different transition time calculated by shooting method.}\label{doublewell4}
 %\label{fig:subfig}%
 \end{figure}

\begin{figure}
\centering
\includegraphics[width=1\textwidth]{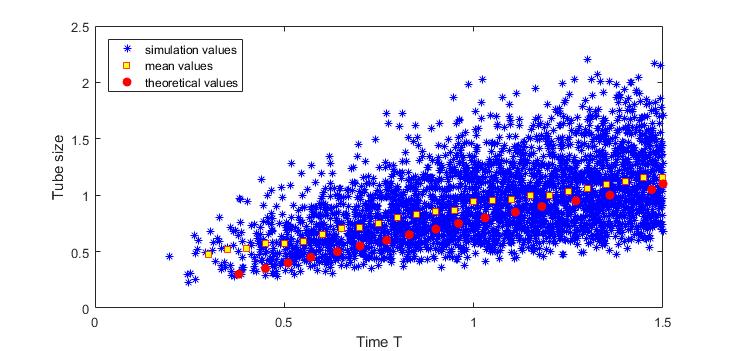}\\
\caption{The transition times and tube sizes of 3165 transition paths.}\label{doublewell5}
\end{figure}

So we restrict our attention in relative short time interval $[0,1.5]$ and the first transition behavior. Figure $\ref{doublewell3}$ shows the graphs of $S^{mOM}_{T}(\psi_T)$. Figure $\ref{doublewell4}$ shows the most probable transition paths for different tube sizes $\delta$ and transition time $T$. These paths are uniformly bounded in domain $[-1,1]$, so in this example we use modified action functional to characterize the transition behavior. Thus the numerical result of the most probable transition times $T^{\delta}_{x_0\rightarrow x_f}$ for different $\delta$ is obtained which are shown in Figure $\ref{doublewell5}$ in red spots.

Furthermore,  we simulated 30000 sample paths starting at -1 and there are 3165 transition paths (the first transition occurred before time 1.5). For every transition path we recorded its transition time $T$ and calculated its corresponding most probable transition path and tube size $\delta$ (i.e. this transition path can be contained in the tube of this MPTP with this tube size).   All the pairs $(T,\delta)$ were draw in Figure $\ref{doublewell5}$ by blue stars. The yellow squares are the mean values of the tube sizes. We separated the time interval [0,1.5] into subintervals $[0,0.25),[0.25,0.3),[0.3,0.35),[0.35,0.4),\cdots,[1.45,1.5]$ and calculated the mean tube size values in every subinterval. As shown in Figure $\ref{doublewell5}$, this numerical simulation shows that our results   characterize the transition behavior.


\newpage

\section{Discussion}\label{conclusion}

We now summarize our work and highlight the differences with relevant works.

\subsection{Our contribution}
Under some mild assumptions, we have   estimated   the  most probable transition time between metatstable states, for stochastic dynamical systems   with  non-vanishing Brownian noise. The problem is represented by a double optimization on the probability that sample paths staying in a tube surrounding the most probable transition pathway.   We have provided estimates for  the most probable transition time.

In our framework, we have adopted the original idea of Onsager-Machlup framework \cite{Machlup1953}, using the concept of tubes surrounding the transition path to study the transition time. Instead of letting the tube size $\delta$ tending to 0, we require this tube size $\delta$ to be a positive constant. This promises the double optimization problem is well-defined. Since when $\delta\rightarrow0$ all tube probabilities are 0. This requirement makes sense as  the noise intensity is non-vanishing,  which is different from the case in Freidlin-Wentzell's large deviation theory \cite{Friedlin1998,Heymann2008}.

Our method can be extended to higher dimensional systems. Similarly, Lemma $\ref{lemma2}$ shows that  the probability that the solution process staying in the tube of a transition path has a power decay law upper bound. It is similar to the one-dimensional case that the most probable transition time has upper and lower bounds. The probability of the solution process staying in the tube of a transition path can be computed approximately in higher dimensional case:
\begin{equation}
\mathcal{P}^{x_0}\{\|X_t-\psi(t)\|_T< \delta\}\approx\exp(-\frac{S^{OM}_T(\psi)}{c^2})\cdot \mathcal{P}^{x_0}\{\|B^c_t-x_0\|_T< \delta\},~\delta\downarrow0.
\end{equation}
  The probability $\mathcal{P}^{x_0}\{\|B^c_t-x_0\|_T< \delta\}$   monotonically decreases  in  $T$. So  we could define a modified OM action functional,  if we have an appropriate analytical estimation for the probability $\mathcal{P}^{x_0}\{\|B^c_t-x_0\|_T< \delta\}$.

  There are some works related to the heuristic discussions and measurements  of transition time. These include    the transition time distribution  \cite{Malinin2010,Carlon2018,Laleman12017,Janakiraman2018} and   the expected transition time     under small noise intensity \cite{Barret2015}.

  %and references for exit phenomena  \cite{Kim2015,Bucher2016} discussed the first passage problem. The exit problem is also an interesting and related topic that has been discussed widely \cite{Ikeda1980,Oksendal2003,Imkeller2006,Imkeller2009,Yang2008}.

\subsection{The difference between   double optimizations of  tube probability and of Onsager-Machlup action functional}

A  \textbf{double optimization problem on the Onsager-Machlup action functional} was investigated in \cite{Du2020}:
\begin{equation}\label{optLD}
\inf_{T>0}\inf_{\psi\in \bar{C}[0,T]} S^{OM}_T(\psi).
\end{equation}
This problem aims to study the property of the OM action functional.
Although \cite{Du2020} and we focus on the same stochastic dynamical systems, our result is different.   Du  et al.  \cite{Du2020} focused on the OM action functional in the $\delta\rightarrow0$ scaling, and the effect of the probability $\mu_{B^c}(K_T(0,\delta))$ is ignored when $T$ varies.

In   the estimation $(\ref{muestimation})$, it can be seen that the OM action functional term   characterizes the geometrical features of a transition path, while  the term
$\mu_{B^c}(K_T(0,\delta))$ characterizes the diffusion ability of the path. So for a stochastic system with  non-vanishing Brownian noise, the most probable transition path and the most probable transition time are supposed to be determined by these two terms (in the original Onsager-Machlup context). Although \cite{Du2020} ignored the effect of $\mu_{B^c}(K_T(0,\delta))$, it provides     valuable insights on  the OM action functional.

\subsection{The difference between our work and the large deviation theory}

Transition phenomena have been treated in the large deviation theory (i.e., under sufficiently small noise). The large deviation theory focuses on the following system \cite{Friedlin1998}:
\begin{equation}\label{ldp}
dX^{\varepsilon}_t=b(X^{\varepsilon}_t)dt+\sqrt{\varepsilon} \; dB_t,~t\geq0,~X_0^{\varepsilon}=x_0\in \mathbf{R}^k,
\end{equation}
where $\varepsilon\rightarrow0$. The Freidlin-Wentzell  (FW)  action functional $S^{FW}_T(\psi)$ is
\begin{equation}
S^{FW}_T(\psi)=
\int_{0}^{T}L^{FW}(\dot{\psi},\psi)dt,
\end{equation}
if $\psi\in C(0,T)$ is absolutely continuous and the integral converges, otherwise denote that $S^{FW}_T(\psi)=\infty$. Here the Lagrangian $L^{FW}(x,y)$ is given by
\begin{equation}
L^{FW}(x,y)=\langle y-b(x),y-b(x)\rangle.
\end{equation}
When the transition time was considered as a factor in transition phenomena, the quasi-potential in \cite{Heymann2008} was defined as
\begin{equation}\label{potential}
\mathbf{V}(x_0,x_f)=\inf_{T>0}\inf_{\psi\in \bar{C}[0,T]} S^{FW}_T(\psi),
\end{equation}
which is used to find the `global' most probable transition path and the corresponding most probable transition time. The large derivation theory asserts that, for $\delta$ and $\varepsilon$ sufficiently small,
\begin{equation}
\mathcal{P}^{x_0}\{\sup_{0\leq t\leq T}|X^{\varepsilon}_t-\psi(t)|\leq \delta\}\approx\exp(-\varepsilon^{-1}S^{FW}_T(\psi)).
\end{equation}
It was shown in \cite{Heymann2008} that the most probable transition time between two metastable states of system ($\ref{ldp}$) is infinite. Since the noise intensity $\sqrt{\varepsilon}$ tends to 0, the behavior of the system is closely  influenced by that of  the deterministic system. Hence the transition between two distinct  metastable states needs infinite time to make it happen most likely.

The modified OM functional in  this small  noise case is
\begin{equation}
\lim_{c\rightarrow0} S^{mOM}_T(\psi)=\lim_{c\rightarrow0}\frac{1}{2}\int_{0}^{T}[(\dot{\psi}-b(\psi))^2+c^2b'(\psi)+\frac{\pi^2 c^4}{4\delta^2}]dt=S^{FW}_T(\psi).
\end{equation}
In particular, the estimation of the most probable transition time $T_{x_0\rightarrow x_f}$ for     Example 1 in  Section $\ref{example}$ is thus \begin{equation}
T^{\delta}_{x_0\rightarrow x_f}=\frac{2\delta|x_f-x_0|}{\pi c^2}.
\end{equation}
When $c\rightarrow0$ then $T^{\delta}_{x_0\rightarrow x_f}\rightarrow\infty$ for any $\delta>0$. This result is consistent with the large deviation theory.

\section*{Acknowledgements}
The authors would like to thank   Jianyu Hu,   Pingyuan Wei,   Xiujun Cheng,   Qiao Huang,   Ao Zhang and Yancai Liu for helpful discussions. This work was partly supported by the  NSFC grants 11531006 and 11771449.


%\begin{acknowledgements}
%\end{acknowledgements}




\end{document}